\documentclass[aps,pra,groupedaddress,floats,reprint]{revtex4-2}

\usepackage[english]{babel}

\usepackage{amsfonts,graphicx,mathrsfs}

\usepackage[colorlinks=true]{hyperref}

\usepackage{epstopdf}
\usepackage{textcomp}
\usepackage{xcolor}
\usepackage{physics}
\usepackage{comment}
\usepackage{amssymb}
\usepackage{bm}

\renewcommand{\vec}[1]{\mathbf{#1}}

\newcommand{\sign}{\text{sign}}

\newcommand{\eu}{\mathrm{e}\mkern1mu}

\begin{document}

\title{Universal Analytic Solution for the Quantum Transport of Structured Matter-Waves in Magnetic Optics}

\author{N.V. Filina}%
\email{nvfilina@bk.ru}%
\affiliation{School of Physics and Engineering,
ITMO University, St. Petersburg, Russia 197101}%

\author{S.S. Baturin}%
\email{s.s.baturin@gmail.com}%
\affiliation{School of Physics and Engineering,
ITMO University, St. Petersburg, Russia 197101}%
\date{\today}

\begin{abstract}

We present a closed-form analytic solution for the propagation of an arbitrary charged scalar state in a non-uniform magnetic field. The dynamics are governed by classical beam optics parameters (Courant-Snyder parameters), the Twiss functions, and phase advance, revealing a direct map between quantum evolution and its classical counterpart. The solution decomposes into three components, exhibiting a complex rotation dependent on the sign of the orbital angular momentum (OAM) projection, alongside an intrinsic distortion from interference governed by a generalized Gouy phase. For a relevant Glaser-type magnetic field and a half-blocked twisted electron, we demonstrate that asymmetry reveals interference-driven dynamics beyond rigid rotation. Our fully relativistic framework provides a practical tool for predicting beam behavior in particle accelerators and electron microscopes.

\end{abstract}


\maketitle


Twisted charged particles, quantum states with helical wavefronts, and a quantized projection of orbital angular momentum (OAM) represent a fundamental class of topological excitations. Their unique properties are used in electron microscopy \cite{Schattschneider:2014,SCHACHINGER2015,Wang2021}, quantum information \cite{PhysRevApplied.11.064058}, and precision material probes \cite{Grillo:2017}. Furthermore, their potential for high-energy physics, enabling novel collision geometries and radiation signatures, is driving significant interest in accelerator-based experiments \cite{IvanovPubl,ct5,ct6,PhysRevD.85.076001}.

The controlled propagation of these structured states in non-uniform magnetic fields is particularly critical for applications in particle accelerators and electron microscopes. Although analytic solutions exist for pure OAM states in simplified fields \cite{Silenko2021,PhysRevA.109.L040201}, the dynamics of general wavepackets, especially the asymmetric or truncated states prevalent in experiments, have until now relied on numerical treatments or restrictive assumptions \cite{ct4,Schattschneider:2014,SCHACHINGER2015}. This gap is especially pertinent for designing and interpreting high-energy tests of twisted particle collisions \cite{PhysRevD.101.096010,PhysRevD.101.016007} and radiation \cite{PhysRevA.109.012222}.

This work presents a universal analytic solution for the quantum propagation of an arbitrary charged scalar state in a nonuniform solenoidal magnetic field. We demonstrate that the complete quantum evolution is uniquely parametrized by classical beam optics quantities: the Twiss $\beta$-function and the phase advance. This mapping, rooted in the Quantum Arnold Transformation approach, reveals a decomposition of the state into components whose complex rotation and interference are not captured by Bohmian trajectory models \cite{Schattschneider:2014,SCHACHINGER2015, Silenko2}.

Consequently, our solution provides a practical tool for modeling beam distortion and Gouy phase evolution in realistic optical systems, including edge effects. It thus offers a unified framework for interpreting twisted particle dynamics in close to experimental conditions.


We consider a set of perfectly cylindrical, on-axis solenoids. Elliptical imperfections can be treated via the perturbation theory approach. Let $z$ be the symmetry axis. Near this axis, one may linearize the magnetic field using the Biot-Savart law (see Appendix \ref{app:1}) and obtain
\begin{align}
\label{eq:B}
    \vec{B}^T = \left[- \frac{x}{2} \partial_z B_z(z) ; - \frac{y}{2} \partial_z B_z(z); B_z(z) \right].
\end{align}
In practice, a discrete solenoid assembly or a measured field map can be represented by a smooth on-axis profile $B_z(z)$; the associated near-axis transverse components then follow from Maxwell's equations via Eq.~\eqref{eq:B}.
This setup is common in electron microscopes and photoinjectors in particle accelerators.

In the Coulomb gauge, the vector potential corresponding to the above magnetic field can be expressed as follows: \cite{Filina2024}
\begin{align}
\label{eq:vpot}
\vec{A}^T = \left[-\frac{y}{2} B_z(z); \frac{x}{2} B_z(z); 0 \right].
\end{align}
We emphasize that vector potential in Eq. \eqref{eq:vpot} represents the unique form mandated by the Coulomb gauge condition $\mathrm{div}\vec{A} = 0$ when applied to our magnetic field in Eq.\eqref{eq:B} (as shown in Appendix \ref{app:3}). We emphasize that within the near-axis expansion of Eq.~\eqref{eq:B}, imposing the Coulomb gauge selects Eq.~\eqref{eq:vpot} (Appendix~\ref{app:3}), yielding a minimal Hamiltonian while keeping transverse (fringe) components self-consistently included.
As a result, edge effects are inherently accounted for in all subsequent analysis.

We begin with Dirac's equation in a stationary magnetic field with minimal coupling
\begin{align}
\left( \bm{\alpha} \hat{\bm{\pi}}+\beta m \right) \Psi=i\partial_t \Psi
\end{align}
with $\hat{\bm{\pi}}=\hat{\vec{p}}-q{\vec{A}}(z)$ being the kinetic momentum operator and $\Psi(\vec{r},t)=\Psi_{st}(\vec{r}) \exp(-iEt)$ - being a bispinor wave function with $E$ - the fixed energy of the state. Above and further, we use the natural unit system with $\hbar = 1$, $c = 1$, and assume that $q$ is the particle charge.

It is known \cite{SilenkoFW,SilenkoG,Silenko2021} that the problem can be reduced to the Pauli-like equation for the upper spinor $\Phi_{FW}$:
\begin{align}
\label{eq:Pauli}
\left(\hat{\pi}^2- q\vec{B}\bm{\sigma}\right)\Phi_{FW}=k^2 \Phi_{FW}.
\end{align}
via the Foldy-Wouthuysen (FW) transformation \cite{FW} with the consequent squaring of the FW Hamiltonian operator. We note that the same result can be achieved without the FW transformation as well \cite{Chuprikov2015}. In the Eq.\eqref{eq:Pauli}, the wave vector $k$ of the state is defined as $k^2=E^2-m^2$.
For electron microscopes and particle accelerators, it is common for the transverse momentum of the particle to be significantly less than the longitudinal momentum, i.e., $p_z \gg p_\perp$, as it is desirable to keep the transverse emittance small. Thus, we seek a solution in the form
\begin{align}
\label{eq:Psi}
    \Phi_{FW}(\vec{r}, t) = \phi_{FW}(\vec{r}) \exp(ikz), 
\end{align}
where we assume that the amplitude $\phi_{FW}$ depends slowly on $z$. Under the paraxial approximation $p_z \gg p_\perp$ we have $\partial^2_z \Phi_{FW} \simeq -k^2 \Phi_{FW} + 2 i k \exp(ikz) \partial_z \phi_{FW}$ and Eq.\eqref{eq:Pauli} transforms to
\begin{align}
\label{FW:par}
\left(\hat{\pi}_\perp^2-q\vec{B}\bm{\sigma}\right)\phi_{FW}=2ik\partial_z \phi_{FW}.
\end{align}
Here $\hat{\bm{\pi}}_\perp= \hat{\vec{p}}_\perp - q \vec{A}_\perp$ - is the transverse part of the kinetic momentum operator. We note that according to the Eq.\eqref{eq:B} at the entrance and exit of any solenoid, there is a transverse magnetic field that, through the Pauli term $q\vec{B}\bm{\sigma}$, couples spin and spatial coordinates as it does not commute with $\hat{\pi}^2$ and explicitly depends on $z$. 
This may result in mixing of spin states and nontrivial spin dynamics. In the present study we omit this interaction and focus only on one of the spinor $\phi_{FW}=(\psi_{u},\psi_{d})^T$ components  $\psi_{u,d}$ assuming the norm of the Pauli operator to be small compared to the norm of the kinetic momentum squared i.e. the ``transverse energy" is significantly greater than the interaction energy of the spin with the magnetic field.  
The final scalar equation that describes the dynamics in the leading order reads
\begin{align}
\label{eq:CKH}
    i \frac{\partial \psi}{\partial z} &= \left[\hat{\mathcal{H}}_{\perp}(z) - \sign(q)  \Omega(z) \hat{L}_z \right] \psi, 
    \nonumber\\
    \hat{\mathcal{H}}_{\perp}(z) &=\frac{\hat{p}_\perp^2}{2} + \frac{\Omega^2(z) \hat{\rho}^2 }{2} .
\end{align}
where we have introduced normalized spatial coordinates $\tilde z = z/(k \rho_H^2)$, $\tilde x= x/\rho_H$ and $\tilde y= y/\rho_H$ and dropped the tilde for clarity; $\hat{L}_z$ is the operator of $z$-projection of the orbital angular momentum and $\hat{\rho}^2=\hat{x}^2+\hat{y}^2$. Magnetic length $\rho_H$ and characteristic frequency are defined as
\begin{align}
\label{eq:param}
\rho_H=\frac{2}{\sqrt{|q|\max{|B(z)|}}},~~~ \Omega(z)=\frac{2B(z)}{\max{|B(z)|}}.
\end{align}

The equation \eqref{eq:CKH}, completed with the initial conditions, can be solved analytically for the Glaser field, free space, and some other special magnetic field configurations. Several approaches allow one to build a solution. The first detailed solution to this problem was derived by Lewis \cite{Lewis1,Lewis2} and later developed in \cite{Lewis}. One recent result that explored this particular equation in the context of twisted-electron propagation is Ref.~\cite{NUF21}. For completeness, Courant--Snyder/Twiss-type parametrization of transverse wave-packet dynamics in uniform axially symmetric magnetic fields were also discussed in Ref.~\cite{KarlovetsNJP2021}; the emphasis there is on packet (second-moment/emittance) evolution rather than on a mode-resolved propagating wavefunction.

The Quantum Arnold Transformation (QAT) \cite{QAT1,QAT2,QAT3} provides an alternative derivation of the analytic solution of Eq.~\eqref{eq:CKH} (equivalently, Eq.~\eqref{eq:HO}) discussed, for example, in Ref.~\cite{NUF21}, but in a form that is more convenient for the present magnetic-optics setup and makes the dynamical structure transparent. Following Ref.~\cite{Filina2023}, we use the Ermakov mapping operator to construct the solution within the QAT framework.

First, we note that, due to the axial symmetry, the following commutator vanishes
\begin{align}
[\hat{\mathcal{H}}_{\perp}(z'),\Omega(z'') \hat{L}_z]=0, \forall z',z''
\end{align}
and we can write 
\begin{align}
\label{eq:psio}
\psi=\exp\left[\sign(q) ~i\int\limits_0^z \Omega(z')dz'  \hat{L}_z \right] \widetilde \psi,
\end{align}
where $\widetilde \psi$ is the solution to the Schr\"{o}dinger type equation of the two-dimensional harmonic oscillator
\begin{align}
\label{eq:HO}
    i \frac{\partial \widetilde{\psi}}{\partial z}= \hat{\mathcal{H}}_{\perp}(z) \widetilde{\psi}
\end{align}
with $\hat{\mathcal{H}}_{\perp}(z)$ given in Eq.\eqref{eq:CKH}.
Let $\psi_{h.o.}$ be a solution to the Eq.\eqref{eq:HO} with $\Omega(z)\equiv1$
\begin{align}
\label{eq:HOz}
i \frac{\partial \psi_{h.o.}}{\partial z} &= \hat{\mathcal{H}}_{h.o.} \psi_{h.o.}, 
    \nonumber\\
    \hat{\mathcal{H}}_{h.o.} &=\frac{\hat{p}_\perp^2}{2}+\frac{\hat{\rho}^2 }{2} .
\end{align}
Then $\widetilde \psi$ solution to the equation \eqref{eq:HO} for an arbitrary $\Omega(z)$ can be written in terms of the Ermakov mapping operator $\hat{\mathcal{E}}$ as 
\begin{align}
\label{eq:erm}
&\widetilde{\psi}(\rho,\phi,z)=\hat{\mathcal{E}}\psi_{h.o.}\\&=\frac{1}{b(z)}\psi_{h.o.}\left[\frac{\rho}{b(z)},\phi,\int\limits_0^z \frac{d\tilde {z}}{b^2(\tilde{z})}\right]\exp\left[i \frac{b'(z)}{b(z)} \frac{\rho^2}{2} \right], \nonumber
\end{align}  
that rescales the solution $\psi_{h.o.}$ of Eq.\eqref{eq:HOz} and modifies the phase.  The prime symbol denotes the total derivative,
and the parameter $b(z)$ must satisfy the Ermakov-Pinney equation \cite{Ermakov, Pinney}:
\begin{align}
\label{eq:EPeq}
    &b'' +\Omega^2(z)b = \frac{1}{b^3}, \\
    &b(0)=b_0,~~b'(0)=b'_0. \nonumber
\end{align}
We note that the ratio $-b'/b$ has the meaning of the inverted radius of curvature of the wave front $1/R(z)$ that is commonly used in Gaussian optics. In accelerator physics and microscopy this equation is known as an envelope equation where $b^2=\beta_t$ has the meaning of the normalized \textit{Twiss $\beta$-function} and $-b'/b=\alpha_t/\beta_t $, where $\alpha_t$ is the \textit{Twiss $\alpha$-function} (see for example Refs.\cite{SYL,Reiser} for the definitions of $\beta_t$ and $\alpha_t$).    

To proceed further, we note that at $z=0$ the wave function $\widetilde \psi (\rho,\phi,0)$ is connected to the wave function $\psi_{h.o.}(\rho,\phi,0)$ as
\begin{align}
\label{eq:hoz}
\psi_{h.o.}(\rho,\phi,0)=b_0 \widetilde\psi(b_0\rho,\phi,0) \exp \left(-i b_0 b'_0 \frac{\rho^2}{2} \right).
\end{align}   

The wave function $\psi_{h.o.}$ is found with the help of an established formula \cite{Landau,Davidov}
\begin{align}
\label{eq:expan}
    \psi_{h.o.}\left(\rho, \phi, z\right) &= \sum\limits_{n,l} c_{n,l} \psi_{n,l}^c \left(\rho, \phi\right) \exp\left(-i \varkappa_{n, l} z\right),
\end{align}
where $\psi_{n,l}^c$ are the eigenstates of the Hamiltonian $\mathcal{H}_{h.o.}$ and the coefficients $c_{n, l}$ are the corresponding overlap integrals
\begin{align}
\label{eq:cnlint}
    c_{n, l} = \int\limits_0^{2 \pi} \int\limits_0^\infty \psi_{h.o.}\left(\rho, \phi, 0\right) \psi^{{c}^{*}}_{n,l} \left(\rho, \phi\right) \rho d\rho d\phi.
\end{align}

The eigenstates $\psi^{{c}}_{n,l}$ are the well-known
Laguerre-Gaussian states 
\begin{equation}
    \begin{gathered}
    \label{eq:psi0}
            \psi_{n, l}^c (\rho, \phi) = N_{n, l} \rho^{|l|} \mathcal{L}_{n}^{|l|}\left(\rho^2\right) \exp\left(-\rho^2/2 + i l \phi\right).
    \end{gathered}
\end{equation}
where $\mathcal{L}_{n}^{|l|}(x)$ are the generalized Laguerre polynomials and the corresponding eigenvalues $\varkappa_{n,l}$ are given by
\begin{align}
    \varkappa_{n,l} = (2n + |l| + 1).
\end{align}
The normalization coefficient $N_{n, l}$ reads
\begin{align}
\label{eq:normc}
    N_{n, l} = \sqrt{\frac{n!}{\pi (n+|l|)!}}.
\end{align}
We notice that we can factor out the rotation operator and rewrite the sum Eq.\eqref{eq:expan} as  
\begin{align}
\label{eq:arcz}
    \psi_{h.o.}\left(\rho, \phi, z\right) = &\exp\left(- i \hat{L}_z z \right) \psi_{+} + \exp\left(i \hat{L}_z z \right)\psi_{-}\nonumber  \\&+\psi_0,
\end{align}
where we have introduced notations
\begin{equation}
\label{eq:psiparts}
    \begin{gathered}
        \psi_{+}(\rho, \phi, z) = \sum\limits_{n, l > 0} c_{n, l} \psi_{n, l}^c (\rho, \phi) \exp\left[ - i (2n+1) z\right], \\
        \psi_{-}(\rho, \phi, z) = \sum\limits_{n, l < 0} c_{n, l} \psi_{n, l}^c (\rho, \phi) \exp\left[ - i (2n+1) z\right], \\
        \psi_{0}(\rho, \phi, z) = \sum\limits_{n} c_{n, 0} \psi_{n, 0}^c (\rho, \phi) \exp\left[ - i (2n+1) z\right]. \\
    \end{gathered}
\end{equation}
Now combining Eq.\eqref{eq:psi0}, Eq.\eqref{eq:erm} and Eq.\eqref{eq:psiparts} we are ready to write down the solution $\psi$ to the initial problem Eq.\eqref{eq:CKH}
\begin{align}
\label{eq:psifull}
    &\psi\left(\rho, \phi, z\right) = \nonumber \\ &\exp\left[i \varphi_+(z) \hat{L}_z \right] \hat{\mathcal{E}}\psi_{+} + \exp\left[i \varphi_{-}(z)\hat{L}_z \right] \hat{\mathcal{E}}\psi_{-}\nonumber  \\&+\exp\left[\sign(q) ~i\int\limits_0^z \Omega(\tilde z)d\tilde z  \hat{L}_z \right]\hat{\mathcal{E}}\psi_0,
\end{align}
where the rotation angles $\varphi_{\pm}$ are given by
\begin{align}
\label{eq:ang}
    \varphi_{\pm}(z) = \int \limits^{z}_0 \left[\sign(q)\;\Omega(\tilde z) \mp \frac{1}{b^2(\tilde z)}\right] d \tilde z.
\end{align}
Above we used the fact that the orbital angular momentum operator $\hat{L}_z$ is the generator of rotations about the $z$-axis and the unitary operator $e^{-i\theta \hat{L}_z}$ implements a rotation by angle $\theta$ via the exponential map from the Lie algebra to the rotation group \cite{warsh}.

Equation \eqref{eq:psifull}, in conjunction with Eq. \eqref{eq:ang}, provides a closed-form analytic solution for the propagation of an arbitrary charged scalar state within a non-uniform magnetic field generated by on-axis solenoids. This solution comprises three distinct components, each undergoing a unique rotation dictated by the mutual orientation of the orbital angular momentum (OAM) projection, the magnetic field direction, and the charge's sign.

The transformation of the initial state is governed entirely by the function $b$, which represents the square root of the normalized \textit{Twiss $\beta$-function}. This directly highlights the profound connection between the quantum state's evolution and its classical counterpart, which is defined solely by the magnetic optics parameters.

The rotation angle for each component has two contributions: the Larmor rotation and a term $\mp \int^z_0 b^{-2} d \tilde{z}$ (for definite positive/negative OAM projections) equivalent to the classical \textit{phase advance}. However, rotation is not the only dynamic effect. As seen in the decomposition of Eq. \eqref{eq:psiparts}, each wavefunction part $\psi_{\pm,0}$ also exhibits intrinsic dynamics governed by the residual phase factor $\exp[-i(2n+1) \int^z_0 b^{-2} d \tilde z]$. Consequently, the inevitable distortion (changing interference pattern that deviates from a simple rigid rotation) of the intensity profile arises from two distinct mechanisms: the relative rotation of the state's components and their intrinsic deformations associated with this residual Gouy phase.

In the special case where $b_0=1$ and the initial state at the focal point ($b'_0=0$) is a pure twisted state $\psi_{n,l}^c$, the general solution of Eq.\eqref{eq:psifull} reduces to a single term. This specific solution
\begin{align}
\label{eq:nns}
&\psi(\rho,\phi,z)=\frac{1}{b(z)}\psi_{n,l}^c\left[\frac{\rho}{b(z)},\phi\right]\exp\left[i \frac{b'(z)}{b(z)} \frac{\rho^2}{2} \right]\times \\ &\exp\left[-i (2n+|l|+1)\int\limits^z_0 \frac{d\tilde z}{b(\tilde z)^2}+\sign(q) ~i l\int\limits_0^z \Omega(\tilde z)d\tilde z\right] \nonumber\end{align}
recovers the results previously reported in Refs. \cite{NUF21, Filina2023}. Here, we identify the factor
\begin{align}
\varphi_G=(2n+|l|+1)\int\limits^z_0 \frac{d\tilde z}{b(\tilde z)^2}
\end{align}
as the generalized Gouy phase. By virtue of the Ermakov operator, this phase is directly mapped to the dynamic phase of $\psi_{n,l}^c$. We therefore conclude that the Gouy phase corresponds precisely to the dynamic phase, where the simple linear dependence on $z$ is generalized and replaced by the \textit{phase advance} of the magnetic optics, $\int^z_0 \frac{d \tilde z}{b(\tilde z)^2}$.

\begin{figure}[t]
    \centering
\includegraphics[width=0.47\textwidth]{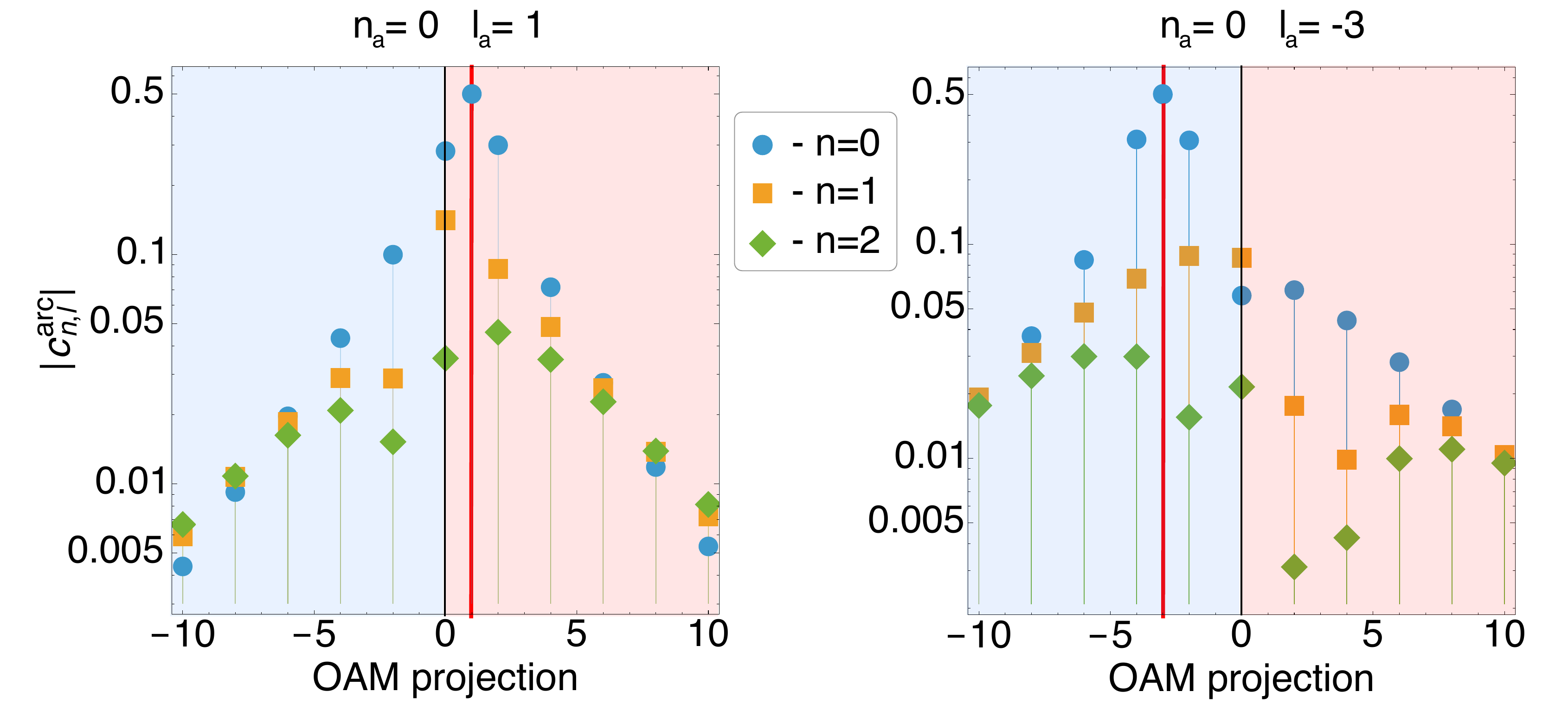}
\caption{Absolute value for the expansion coefficients $c_{n,l}^{arc}$ for the half-blocked twisted state for two initial values of the OAM projection $l_a=1$ - left panel and $l_a=-3$ - right panel. Radial number is zero in both cases $n_a=0$. The plots are on a logarithmic scale.}
\label{fig:Fig1}
\end{figure}

\begin{figure*}[t]
    \centering
\includegraphics[width=1\textwidth]{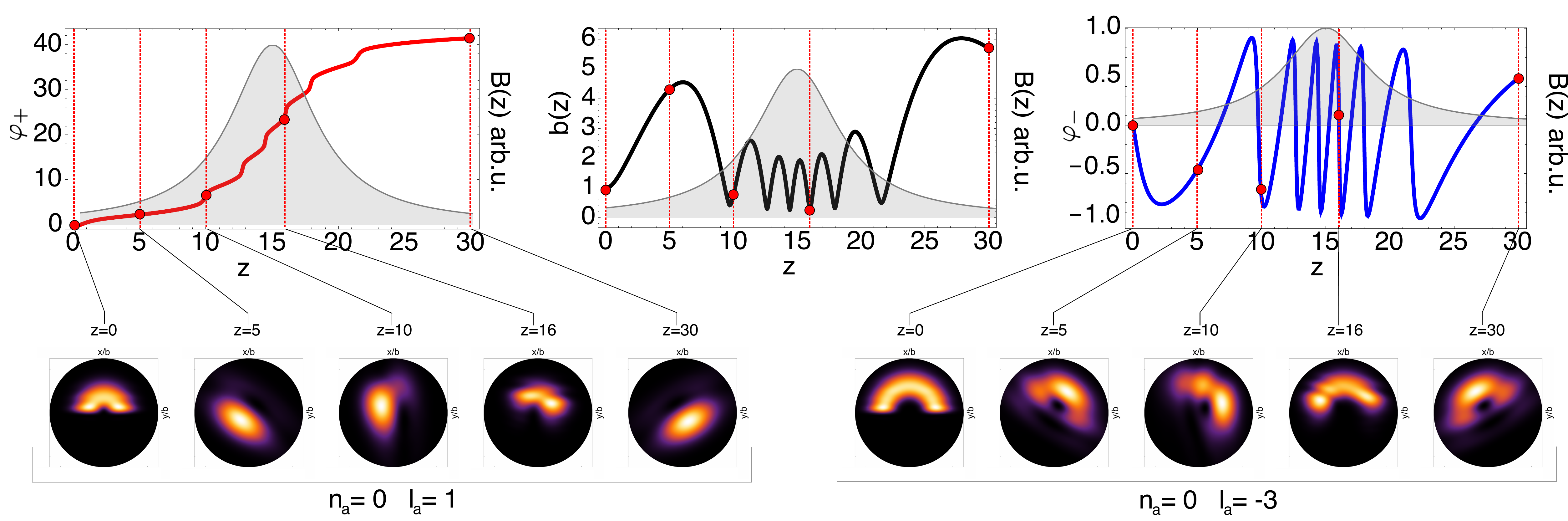}
\caption{Rotation angles $\varphi_{\pm}(z)$ calculated with the help of the Eq.\eqref{eq:ang} for the case of the Glaser field Eq.\eqref{eq:GZF} with the parameters $a=4$ and $c=15$ (upper left and upper right panels). Parameter $b(z)$ (upper middle panel) calculated based on the analytical expressions \cite{Lewis,NUF21} of the solution to the Eq.\eqref{eq:EPeq} with the Glaser field Eq.\eqref{eq:GZF} and $b_0=1$, $b_0'=0$. Intensity distributions (lower panel) $|\psi_{arc}|^2$ of the half-blocked twisted state Eq.\eqref{eq:psiarc} for different positions $z$ inside the Glaser field calculated with the help of the Eq.\eqref{eq:psifull}, Eq.\eqref{eq:psiparts} and Eq.\eqref{eq:cnl}.}
\label{fig:Fig2}
\end{figure*}

The structure of the solution in Eq.\eqref{eq:psifull} explicitly shows that if the decomposition of the initial state over the eigenbasis \eqref{eq:cnlint} includes both projections of the OAM (a typical experimental scenario), then the transformation of the state along the $z$ axis is not just a rotation, but an interplay between three parts. Thus, the Bohmian trajectory interpretation that was applied to describe the propagation of the non-pure highly asymmetric twisted states \cite{Schattschneider:2014,SCHACHINGER2015,Silenko2} is limited due to strong interference effects. Moreover, if the state is not matched to magnetic optics, then the experimental observation of a Landau-Zeeman-Gouy phase suggested in Ref.\cite{Bliokh2012} is obscured by the intrinsic deformations determined by the residual dynamic phase.   
Indeed, let's consider a half-blocked twisted electron state that was extensively studied in Refs.\cite{Schattschneider:2014,SCHACHINGER2015,Silenko2} as a test bed to reveal the internal motion of the twisted state:
\begin{align}
\label{eq:psiarc}
 \psi_{arc}\left(\rho, \phi, 0\right) = \psi_{n_a, l_a}^c (\rho, \phi) \left[\theta(\phi) \theta(\pi-\phi)\right].
\end{align}

Above, $n_a$ and $l_a$ are the quantum numbers of the twisted state before blocking.
For simplicity, we assume that $b_0=1$ and $b'_0=0$ in Eq.\eqref{eq:hoz} (this does not violate the generality of the consideration). Under this assumption the coefficients $c_{n, l}$ can be calculated as
\begin{align}
\label{eq:cnlint2}
    c^{arc}_{n, l} = \int\limits_0^{\pi} \int\limits_0^\infty\psi_{n_a, l_a}^c (\rho, \phi)  \psi^{{c}^{*}}_{n,l} \left(\rho, \phi\right) \rho d\rho d\phi.
\end{align}
These integrals can be evaluated analytically, exact expression and the calculation procedure can be found in Appendix \ref{app:2}.

In Fig.\ref{fig:Fig1} we plot these coefficients for an initial twisted state with radial quantum number $n_a=0$ and two cases of OAM projections: $l_a=1$ and $l_a=-3$. As anticipated, the peak amplitude occurs at $l = l_a$. However, due to edge effects, the decomposition spectrum is broad and includes amplitudes with OAM projections of both the same and the opposite sign. This latter point demonstrates that the rotation of an asymmetric state is a significantly more complex process than that of a pure twisted state.

While Eq.\eqref{eq:psifull} confirms the proposed structure of the intrinsic ``rotation" for a pure state as discussed in Ref.\cite{Bliokh2012}. However, these rotations cannot be observed in isolation. It either requires a mixture of two or more states to form an interference pattern as suggested in Ref.\cite{Bliokh2012} or requires breaking the axial symmetry as was done in Refs.\cite{Schattschneider:2014,SCHACHINGER2015}. The later triggers additional transformations and results in the change of the shape of the intensity pattern.

 To illustrate this effect, we consider the propagation of the half-blocked twisted state through the Glaser field that has the form \cite{Glazer} 
\begin{align}
\label{eq:GZF}
B_G(z)=B_0 \frac{a^2}{a^2+(z-c)^2}.
\end{align}

With Eq.\eqref{eq:GZF} Ermakov-Pinney equation \eqref{eq:EPeq} can be solved analytically (see Refs.\cite{Lewis,NUF21} for details). For illustration, we adopt Glaser field parameters of $a=4$, $c=15$; the resulting function $b(z)$ is plotted in the upper middle panel of Fig.\ref{fig:Fig2}. Consistent with previous findings \cite{NUF21, Silenko2, Filina2023}, we observe oscillations in $b(z)$ within the magnetic field region where the Glaser field strength is significant, leading to corresponding oscillations in the quantum state dispersion $\sigma_\rho$.

The rotation angles computed from Eq.\eqref{eq:ang} are displayed in the top left and right panels of Fig.\ref{fig:Fig2}. For the $\psi_+$ component, where the OAM projection is aligned with the magnetic field, the result is a steady net rotation. In contrast, the $\psi_{-}$ component exhibits an oscillating rotation angle, producing a characteristic wobbling motion. We note that using one of the angles from Eq. \eqref{eq:ang} is an approximation for the non-pure state. The decomposition coefficients for the half-blocked state span both positive and negative values. Thus, we define the approximate observable net rotation angle as the angle of the group with the largest total intensity, which is the sum of the squared moduli of the decomposition coefficients.  

To analyze the dynamics, we examine two specific cases of half-blocked states: one with $n_a=0$, $l_a=1$ (Fig.\ref{fig:Fig2}, lower left row) and another with $n_a=0$, $l_a=-3$ (Fig.\ref{fig:Fig2}, lower right row). The intensity distribution $|\psi_{arc}|^2$ at various positions $z$ is calculated using Eq.\eqref{eq:psifull}, Eq.\eqref{eq:psiparts}, and the exact coefficients $c_{n,l}$ from Eq.\eqref{eq:cnl}, assuming $b_0=1$ and $b'_0=0$ for simplicity. 
 
 Our analysis reveals that alongside the rotational phenomenon, a significant distortion (changing interference pattern that deviates from a simple rigid rotation) of the intensity pattern occurs - particularly for the $l_a=1$ case - which depends on the phase advance. Notably, when the phase advance reaches values of $\int^z_0 b^{-2} d \tilde z = m \pi, ~~m\in \mathcal{N}$, the structures of the $\psi_+$, $\psi_{-}$, and $\psi_0$ resembles the series at $z=0$. This occurs because the residual Gouy phase factor becomes $\exp[-i(2n+1) m \pi]=(-1)^m$, which does not affect the intensity $|\psi|^2$. At these specific points, the intrinsic dynamic deformation vanishes, leaving only the distortion caused by the differential rotations of the components. 
 
In summary, we have derived a rigorous solution describing the propagation of a complex quantum state in an inhomogeneous magnetic field. Our model accurately represents setups common to electron microscopes and particle accelerators, thereby enabling a detailed analysis of the transport and distortion of twisted charged particles through magnetic optics.

We emphasize that even a slight deformation of an initial pure twisted state, such as a simple asymmetric stretch, introduces multiple terms into its decomposition Eq.\eqref{eq:expan}. This, in turn, triggers complex dynamics: the residual Gouy phase induces non-trivial interference between these components, leading to significant distortion of the propagating state. This distortion can critically affect key observables, including the mean value of the OAM projection, with important implications for detecting the signatures of twisted particles in radiation and high-energy reactions.

We have demonstrated that, despite its complexity, the propagation of a quantum state through an optical system can be completely characterized by the well-established parameters of magnetic optics: the \textit{Twiss $\beta$-function}, the \textit{Twiss $\alpha$-function}, and the \textit{phase advance}. Our description reveals two primary observable effects: a distortion arising from the residual Gouy (dynamic) phase and the simultaneous yet distinct rotation of the three state components corresponding to positive, negative, and zero OAM projections. Notably, the distortion caused by the residual Gouy phase vanishes at points where the phase advance is an integer multiple of $\pi$. We also note that the $z$-dependent basis given in Eq. \eqref{eq:nns} is a natural basis for the system of solenoids because it is a set of eigenvectors for the quadratic self-adjoint Lewis invariant \cite{Lewis} for this magnetic system.

This method provides a straightforward and powerful tool for calculating and analyzing the transport of twisted states in particle accelerators and electron microscopes. It establishes a unified framework for describing the propagation of diverse structured matter waves \cite{Voloch-Bloch:2013,Goutsoulas2021}. A key advantage of this approach is its rigorous relativistic foundation, which enables the accurate modeling of realistic magnetic field configurations and offers a comprehensive description of edge effects at the boundaries of axisymmetric coils.

\begin{acknowledgments}

N.V.F. thanks Dmitry Vagin for the help with the programming. The work was supported by the Priority 2030 Academic Program.

\end{acknowledgments}

\appendix


\section{Linearized magnetic field of a solenoid with a circular cross-section \label{app:1}}


We parametrize the surface of the solenoid by a vector with coordinates $\left[R(z_0) \cos{\varphi}, R(z_0) \sin{\varphi}, z_0 \right]$.
The radius vector from the point $(x, y, z)$ to the point on the surface of the solenoid is given by
\begin{align}
    \vec{r} = \left(
    \begin{matrix}
        x - R \cos{\phi} \\
        y - R \sin{\phi} \\
        z - z_0
    \end{matrix} \right).
\end{align}
The element of the contour $\vec{dl}$ with the current on the surface of the solenoid can be written as
\begin{align}
    \vec{dl} = \left(
    \begin{matrix}
        - R \sin{\phi} \\
        R\cos{\phi} \\
        0
    \end{matrix} \right) d\phi.
\end{align}
Above and further, we drop the argument of R and I for brevity.  
The corresponding vector product reads 
\begin{align}
    \left[ \vec{dl} \times \vec{r} \right] = \left[
    \begin{matrix}
        (z - z_0) R\cos{\phi} \\
        (z - z_0) R\sin{\phi} \\
        - y R \sin{\phi} - x R \cos{\phi} + R^2    \end{matrix} \right] d\phi.
\end{align}
\begin{widetext}
    
Now we are ready to write down the Biot--Savart law. We start from the $x$ component of the magnetic field that has the form
\begin{align}
    B_x = \frac{\mu_0}{4 \pi} \int\limits_{-\infty}^{+\infty}
    \int\limits_0^{2\pi} \frac{(z - z_0) I R \cos{\phi} \; d\phi}{\left[ x^2 + y^2 + (z - z_0)^2 - 2 x R \cos{\phi} - 2 y R \sin{\phi} + R^2 \right]^{3/2}} dz_0.
\end{align}
We consider the magnetic field near the solenoid axis, neglecting the terms of the order $x^2$ and $y^2$ and get
\begin{align}
    B_x = \frac{\mu_0}{4 \pi} \int\limits_{-\infty}^{+\infty}
    \int\limits_0^{2\pi} \frac{(z - z_0) I R \cos{\phi} \; d\phi}{\left[ (z - z_0)^2 + R^2  \right]^{3/2}} \left[ 1 + \frac{3}{2} \frac{2 x R \cos{\phi} + 2 y R \sin{\phi}}{(z - z_0)^2 + R^2} \right] dz_0.
\end{align}
\end{widetext}

Two of the three integrals are zero due to the integrand's periodicity. 
\begin{align}
    B_x = \frac{\mu_0}{4 \pi} \int\limits_0^{2\pi} \int\limits_{-\infty}^{+\infty} \frac{3 I R^2 x \cos^2{\phi} \; (z - z_0) dz_0}{\left[ (z - z_0)^2 + R^2 \right]^{5/2}} d\phi.
\end{align}
It's easy now to evaluate the integral over $\phi$. The result is
\begin{align}
    B_x = x \frac{\mu_0}{4} \int\limits_{-\infty}^{+\infty} \frac{3 I(z_0) R^2(z_0) (z - z_0) dz_0}{\left[ (z - z_0)^2 + R^2(z_0) \right]^{5/2}}.
\end{align}
Next, we calculate the $y$-component of the magnetic field
\begin{align*}
    B_y = \frac{\mu_0}{4 \pi} \int\limits_0^{2\pi} \int\limits_{-\infty}^{+\infty} \frac{3 I R^2 y \sin^2{\phi} \; (z - z_0) dz_0}{\left[ (z - z_0)^2 + R^2 \right]^{5/2}} d\phi.
\end{align*}
Linearization and integration give
\begin{align}
    B_y = y \frac{\mu_0}{4} \int\limits_{-\infty}^{+\infty} \frac{3 I(z_0) R^2(z_0) (z - z_0) dz_0}{\left[ (z - z_0)^2 + R^2(z_0) \right]^{5/2}}.
\end{align}
Finally, we evaluate $B_z$. After the linearization, we get
\begin{align}
    B_z = \frac{\mu_0}{4 \pi} \int\limits_0^{2\pi} \int\limits_{-\infty}^{+\infty} \frac{I R^2 dz_0}{\left[ (z - z_0)^2 + R^2 \right]^{3/2}} d\phi.
\end{align}
Integrating over $\phi$ we obtain
\begin{align}
    B_z = \frac{\mu_0}{2} \int\limits_{-\infty}^{+\infty} \frac{I(z_0) R^2(z_0) dz_0}{\left[ (z - z_0)^2 + R^2(z_0) \right]^{3/2}}.
\end{align}
Now, we can express each component of the magnetic field through the $B_z$. The final form of the linearized magnetic field near the $z$-axis in full accordance with the Biot-Savart law is
\begin{align}
    \vec{B}^T = \left[- \frac{x}{2} \partial_z B_z(z) ; - \frac{y}{2} \partial_z B_z(z); B_z(z) \right].
\end{align}
Therefore, in the case of any axially symmetric electron current distribution, the transverse components of the magnetic field are always non-zero and are proportional to the $z$ derivative of $B_z$.


\section{Calculation of expansion coefficients for the half-blocked twisted state \label{app:2}}


We start from the expression for the scalar product of $\psi_{arc}$ with the eigenfunction $\psi^{c}$ for the case of $b_0=1$ and $b_0'=0$ with Eq.\eqref{eq:psiarc} and Eq.\eqref{eq:hoz} we have 
\begin{align}
\label{eq:cnlintapp}
    c^{arc}_{n, l} = \int\limits_0^{\pi} \int\limits_0^\infty\psi_{n_a, l_a}^c (\rho, \phi)  \psi^{{c}^{*}}_{n,l} \left(\rho, \phi\right) \rho d\rho d\phi.
\end{align}
Using the exact expression for the eigenstate $\psi_{n,l}^c$ given by Eq.\eqref{eq:psi0} we carry out the integration over the angle and get
\begin{align}
&c^{arc}_{n, l}= \\ \nonumber &N_{n,l} N_{n_a,l_a}I_\rho\times\begin{cases}
    \pi , \quad l = l_a \\
    0, \quad l \ne l_a \; \text{and} \; (l-l_a) \; \text{is even} \\
    \frac{2 i}{l_a - l}, \quad (l-l_a) \; \text{is odd}
    \end{cases},
\end{align}
where $N_{n,l}$ and $N_{n_a,l_a}$ are the normalization coefficients given by Eq.\eqref{eq:normc}, and we have introduced the notation for the radial integral
\begin{align}
\label{eq:I2}
    I_\rho = \int\limits_0^\infty \rho^{|l|+|l_a|+1} \mathcal{L}_{n}^{|l|}\left(\rho^2\right) \mathcal{L}_{n_a}^{|l_a|}\left(\rho^2\right) e^{-\rho^2} d\rho. 
\end{align}
The radial integral can be simplified further to 
\begin{align}
I_\rho=\frac{1}{2} \int\limits_0^\infty \xi^\frac{|l|+|l_a|}{2} \mathcal{L}_{n}^{|l|}\left(\xi\right) \mathcal{L}_{n_a}^{|l_a|}\left(\xi\right) \eu^{-\xi} d\xi.
\end{align}
Above we made a substitution $\xi=\rho^2$.
A similar integral can be found in the book of integrals by Gradshteyn and Ryzhik \cite{GR} (7.414 (9)) that is
\begin{align}
\label{eq:GR}
    &\int\limits_0^\infty \eu^{-x} x^{\alpha+\beta} \mathcal{L}_{m}^{\alpha}(x) \mathcal{L}_{n}^{\beta}(x) dx = \nonumber \\ &(-1)^{m+n} \binom{m+\alpha}{n} \binom{n+\beta}{m} \Gamma(\alpha+\beta+1),
\end{align}
where $\binom{m}{n}$ is the binomial coefficient and $\Gamma(\alpha)$ is the Euler Gamma function.
In our case, the only difference is in the degree of $\xi$. To fix this, we use expansion for Laguerre polynomials from \cite{GR} (8.974 (2)) in the form
\begin{align}
\label{eq:wiki}
    \mathcal{L}_{m}^{\alpha}(x) = \sum\limits_{k=0}^m \binom{\alpha-\gamma+m-k-1}{m-k} \mathcal{L}_{k}^{\gamma}(x),
\end{align}
where we take the parameters $\alpha$ and $\gamma$ as
\begin{align}
    \alpha = |l|, \quad \gamma = \frac{|l|-|l_a|}{2}.
\end{align}
\begin{widetext}
So after substitutions \eqref{eq:GR} and \eqref{eq:wiki} in \eqref{eq:I2} we get
\begin{align}
\label{eq:I2f}
    I_\rho =\frac{1}{2} \sum\limits_{k=0}^n \binom{|l|-\frac{|l|-|l_a|}{2}+n-k-1}{n-k} (-1)^{n_a+k} \binom{n_a+|l_a|}{k} \binom{k+\frac{|l|-|l_a|}{2}}{n_a} \Gamma\left(|l_a|+\frac{|l|-|l_a|}{2}+1\right).
\end{align}
The final answer for expansion coefficients is
\begin{align}
   \label{eq:cnl}
    &c^{arc}_{n,l} = \frac{1}{2 \pi} \sqrt{\frac{n! \; n_a!}{(n+|l|)!(n_a+|l_a|)!}}\Gamma\left(\frac{|l|+|l_a|}{2}+1\right)\times \nonumber \\ &\sum\limits_{k=0}^n (-1)^{n_a+k} \binom{n_a+|l_a|}{k} \binom{k+\frac{|l|-|l_a|}{2}}{n_a} \binom{\frac{|l|+|l_a|}{2}+n-k-1}{n-k}    
  \begin{cases}
    \pi , \quad l = l_a \\
    0, \quad l \ne l_a \; \text{and} \; (l-l_a) \; \text{is even} \\
    \frac{2 i}{l_a - l}, \quad (l-l_a) \; \text{is odd}
    \end{cases}.
\end{align}
\end{widetext}

\section{Gauge choice \label{app:3}}
We start with the linearized magnetic field
\begin{align}
\label{eq:appB}
    \vec{B}^T = \left[- \frac{x}{2} \partial_z B_z(z) ; - \frac{y}{2} \partial_z B_z(z); B_z(z) \right].
\end{align}

Following the connection $\mathrm{rot} \vec{A} = \vec{B}$, one may derive a general form for the vector potential that corresponds to this field:
\begin{align}
\label{eq:vpgen}
    \vec{A}=\left(\begin{matrix}  B_z(z)\left[a x + (b-\frac{1}{2}) y \right] \\  B_z(z)\left[(b+\frac{1}{2}) x + c y \right] \\ \partial_z B_z(z)\left[\frac{a}{2} x^2 + b x y +\frac{c}{2} y^2\right] \end{matrix}\right),
\end{align}
Above arbitrary constants $a,b,c$ reflect the gauge freedom.
We depart from the Pauli-like equation for the upper spinor $\Phi_{FW}$:
\begin{align}
\label{eq:appPauli}
\left(\hat{\pi}^2- q\vec{B}\bm{\sigma}\right)\Phi_{FW}=k^2 \Phi_{FW},
\end{align}
and proceed with the paraxial approximation as given below
\begin{align}
\label{eq:appPsi}
    \Phi_{FW}(\vec{r}, t) = \phi_{FW}(\vec{r}) \exp(ikz).
\end{align}
Here we assume that the amplitude $\phi_{FW}$ depends slowly on $z$ thus the following connection $\partial^2_z \Phi_{FW} \simeq -k^2 \Phi_{FW} + 2 i k \exp(ikz) \partial_z \phi_{FW}$ holds. Eq.\eqref{eq:appPauli} with the vector potential in the form \eqref{eq:vpgen} transforms to
\begin{align}
\label{eq:appgeneral}
&\Bigg\{\hat{\pi}_\perp^2-q\vec{B}\bm{\sigma} + i q \partial^2_{zz} B_z(z) \left(\frac{a}{2} x^2 + b x y +\frac{c}{2} y^2\right) - \nonumber\\
&- 2 q \partial_z B_z(z)\left(\frac{a}{2} x^2 + b x y +\frac{c}{2} y^2\right) \hat{p}_z + \nonumber\\
&+ q^2 [\partial_z B_z(z)]^2 \left(\frac{a}{2} x^2 + b x y +\frac{c}{2} y^2 \right)^2 \Bigg\}\phi_{FW} = \nonumber\\
&= 2ik\partial_z \phi_{FW},
\end{align}
where $\hat{\bm{\pi}}_\perp= \hat{\vec{p}}_\perp - q \vec{A}_\perp$ - is the transverse part of the kinetic momentum operator.
To simplify the equation, we should fix the gauge. Due to the general gauge invariance the solution is gauge independent and it is a matter of convenience. It is clear that arbitrary chosen gauge complicates calculations. For instance if we are to chose the Laundau gauge, we get $a=c=0, b=\pm 1/2$ and Eq.\eqref{eq:appgeneral} for $b=+1/2$ reduces to
\begin{align}
&\Bigg\{\hat{\pi}_\perp^2-q\vec{B}\bm{\sigma} + \frac{i}{2} q \partial^2_{zz} B_z(z) x y + i q \partial_z B_z(z) x y \partial_z + \nonumber\\
&+ \frac{1}{4} q^2 [\partial_z B_z(z)]^2 x^2 y^2 \Bigg\}\phi_{FW} = 2ik\partial_z \phi_{FW}.
\end{align}
We note mixed term where the transverse coordinates are multiplied by the $z$-component of momentum. This term does not permit us to factor out the longitudinal and transverse dynamics and complicates further analysis. 

Interestingly, if we impose Coulomb gauge $\mathrm{div}\vec{A}=0$ from Eq.\eqref{eq:vpgen} we get
\begin{align}
B_z(z)\left[a+c\right]+\partial^2_{z}B_z(z)\left[a\frac{x^2}{2}+bxy+c\frac{y^2}{2} \right]=0.
\end{align}
Equation above must be valid for any $x$ and $y$ as well as for any $B_z(z)$, this immediately results in $a=b=c=0$ and giving
\begin{align}
\vec{A}^T = \left[-\frac{y}{2} B_z(z); \frac{x}{2} B_z(z); 0 \right].
\end{align}
and Eq.\eqref{eq:appgeneral} simplifies to Eq.\eqref{FW:par}.


\bibliography{references}


\end{document}